\documentclass[pdflatex,sn-mathphys-num]{sn-jnl}

\usepackage{graphicx}%
\usepackage{subcaption}
\usepackage{multirow}%
\usepackage{amsmath,amssymb,amsfonts}%
\usepackage{amsthm}%
\usepackage{mathrsfs}%
\usepackage[title]{appendix}%
\usepackage{xcolor}%
\usepackage{textcomp}%
\usepackage{manyfoot}%
\usepackage{booktabs}%
\usepackage{algorithm}%
\usepackage{algorithmicx}%
\usepackage{algpseudocode}%
\usepackage{listings}%
\usepackage{braket}
\usepackage{tikz}
\usepackage[european]{circuitikz}
\usetikzlibrary{decorations.pathreplacing}
\usetikzlibrary{shapes.symbols}
\usepackage{physics}

\def\e{\ensuremath{\mathrm{e}}}
\def\i{\ensuremath{\mathrm{i}}}
\def\d{\ensuremath{\mathrm{d}}}
\def\Tr{\ensuremath{\mathrm{Tr}}}

\begin{document}

\title[Article Title]{Heat flow through the quantum heat valve coupled to ohmic baths via a master equation approach}

\author*[1,2]{\fnm{Antti} \sur{Vaaranta}}\email{antti.vaaranta@helsinki.fi}
\author[1]{\fnm{Marco} \sur{Cattaneo}}\email{marco.cattaneo@helsinki.fi}
\author[3]{\fnm{Paolo} \sur{Muratore-Ginanneschi}}\email{paolo.muratore-ginanneschi@helsinki.fi}
\author[2]{\fnm{Jukka} \sur{Pekola} \email{jukka.pekola@aalto.fi}}

\affil*[1]{\orgdiv{Department of Physics}, \orgname{University of Helsinki}, \orgaddress{\street{P.O. Box 43}, \postcode{FI-00014}, \city{Helsinki}, \country{Finland}}}
\affil[2]{\orgdiv{Department of Applied Physics}, \orgname{Aalto University}, \orgaddress{\street{P.O. Box 15100}, \postcode{FI-00076 Aalto}, \city{Espoo}, \country{Finland}}}
\affil[3]{\orgdiv{Department of Mathematics and Statistics}, \orgname{University of Helsinki}, \orgaddress{\street{P.O. Box 43}, \postcode{FI-00014}, \city{Helsinki}, \country{Finland}}}

\abstract{
  We provide a theoretical model for the non-equilibrium steady state heat flow through a quantum heat valve. The model is based on a master equation approach, where the partial secular approximation has been carefully performed in order to obtain accurate results. Our study assumes an ohmic spectral density for the two thermal baths of the model. This is in contrast with previous treatments of the quantum heat valve, where the baths have been assumed as being structured with a peaked spectral density near the resonance frequency of the resonator. These studies have also taken the resonator to be a part of the open quantum system of interest, which results in double counting of the resonator, as the latter appears both in the spectral density of the bath and as a part of the open system. Although this model accounts for the observations in a satisfactory way, it raises issues regarding its physical interpretation. Our method solves this conceptual problem. We apply it to describe an experiment on a quantum heat valve, showing that it successfully captures the experimental results and improves upon the previous theoretical model, which suffered from the resonator double-counting issue.  Our findings confirm that the careful application of the master equation approach, in particular when it comes to the secular approximation, is a useful tool for explaining realistic experimental setups.
}


\maketitle

\section{Introduction}\label{sec:introduction}

The field of quantum thermodynamics \cite{Potts2024,Kosloff2013,Vinjanampathy2016} advances hand-in-hand by theoretical and experimental studies. Theoretical proposals are often followed by experimental realizations, or vice versa, and over time both sides of the same coin are polished, making our understanding on how thermodynamics arises from the quantum phenomena, and the field in general, clearer. The progress in the field is pushed forward for example via the experimental realization of quantum heat engines and refrigerators \cite{Cangemi2024,Bhattacharjee2021}, rectifiers \cite{Pekola2021,Santiago-Garcia2025,Senior2020} and heat valves \cite{Ronzani2018,Xu2021} together with advancing our theoretical understanding of these devices. Each of these proof-of-concept experiments have paved the way towards greater understanding on how thermodynamics works in the quantum realm and offered us tools to manage and exploit the thermal effects at the quantum level.

An example of these tools is the quantum heat valve, which is a tunable element able to modulate heat flow between two thermal reservoirs. In the quantum heat valve, as implemented in Ref.~\cite{Ronzani2018}, a resonator-qubit-resonator chain based on the circuit quantum electrodynamics (cQED) architecture \cite{Vool2017,Rasmussen2021,Krantz2019,Blais2021} is installed between two baths of unequal temperatures. The qubit is of the transmon type \cite{Koch2007}, and its resonance frequency can be varied by applying external magnetic flux to the circuit. This tunable resonance frequency of the qubit directly translates into stronger or weaker heat flow when the system is either fully in resonance or detuned, respectively.


Experimental and theoretical advancements are two sides of the same coin. In the case of the quantum heat valve, the theoretical side was partially studied already in \cite{Karimi2016} in the case of very weak qubit-resonator coupling compared to the resonator decay rate to the environment. The coin was then flipped to the experimental side in Ref.~\cite{Ronzani2018}, where the model from \cite{Karimi2016} was numerically fitted to explain part of the observed data. For the other case, where the qubit-resonator coupling is stronger than the resonator decay rate, the authors of Ref.~\cite{Ronzani2018} made their own phenomenological model based on Fermi's golden rule and succesfully fitted this model to explain the experimental data.

The theoretical side of the coin was polished further in Ref.~\cite{Xu2021}, where the authors gave a rigorous and exact solution to the non-equilibrium steady state heat flow through the quantum heat valve setting using the numerical hierarchical equations of motion (HEOM) approach \cite{tanimura1989time}. They showed that by using HEOM they can obtain a good fit to the experimental data using several different bath spectral densities, as well as obtain qualitative understanding on how the parameters affect the system's behaviour.

In this paper we show that well-matching solutions to experimental data for the heat flow through the quantum heat valve in the non-equilibrium steady state can also be obtained by using the numerically less intensive master equation approach \cite{breuer2002theory} with a careful application of the partial secular approximation \cite{Vaaranta2026}. We call this method the global master equation with partial secular approximation (ME PSA) approach \cite{Cattaneo2019}. We also explore a related approach, namely the unified master equation \cite{Trushechkin2021} that is guaranteed to be completely positive, and find similar results as for the ME PSA approach. 

In addition, we comment on how the inclusion of the resonator within the baths (by modifying the form of the bath spectral density) while still retaining the resonator as part of the system leads into a \textit{double-counting} of the resonator, which is an unphysical assumption. Therefore, in our model we keep the baths ohmic and weakly coupled to the resonators, and show how this approach successfully captures the desired heat flow behaviour with respect to the external magnetic field applied to the qubit.

In this respect, we argue that the careful implementation of the perturbative master equation approach, especially when dealing with the secular approximation, is able to capture the physics of the steady state heat flow. Additionally, it retains a clear physical interpretation of the different model parameters at play.

In \cite{Xu2021} the authors state  that the perturbative master equation approach will certainly fail in quantitative understanding of the experimental results. They were referring to the conventional master equation in full secular equation, whose decay rates can be computed through Fermi's golden rule. It is indeed true that this kind of master equation cannot fully capture the steady-state heat flow due to the incorrect full secular approximation that may remove slowly-rotating terms. Moreover, any perturbative master equation approach for the qubit alone, where the two resonators are considered as part of the baths, cannot describe any scenarios where the qubit-resonator coupling is not weak. This said, we argue that careful treatments based on a perturbative master equation can still capture the steady state of the resonator-qubit-resonator open system. Indeed, in our case, while the ME PSA approach is perturbative and relies on the assumption of weak coupling between the system and the reservoirs, it is still able to describe the behavior of the experimental realization of the quantum heat valve in Ref.~\cite{Ronzani2018}, while at the same time being intuitive and computationally rather lightweight.

This paper is organized as follows: in Section \ref{sec:quantum-heat-valve} we introduce the quantum heat valve in more detail and discuss the modelling of the heat valve both phenomenologically and with the HEOM approach. We also discuss the double-counting issue of the resonators, where the resonator is taken to be both part of the system and effectively part of the bath. We present our solution to this problem, based on the master equation formalism and without using the HEOM method, in Section \ref{sec:results}. Finally, Section \ref{sec:conclusion} draws some concluding remarks.

\section{The quantum heat valve}\label{sec:quantum-heat-valve}

\subsection{Description and open-system formulation}

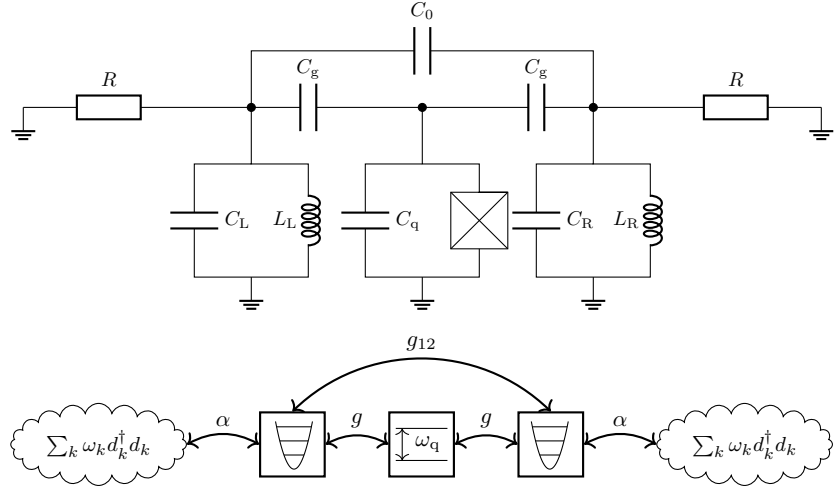
\begin{figure}
  \centering
  \begin{subfigure}{0.75\linewidth}
    \begin{circuitikz}[cute inductors, scale=0.75, transform shape]
      
      \draw (0,0) node[ground] {} to[R=$R$] (3,0) to[C=$C_\text{g}$] (7,0) to[C=$C_\text{g}$] (11,0) to[R=$R$] (14,0) node[ground] {};

      \draw (4,0) -- (4,-1) -- (5,-1) to[L, l_=$L_\text{L}$] (5,-3) -- (4,-3) node[ground] {} -- (3,-3) to[C, l_=$C_\text{L}$] (3,-1) -- (4,-1);

      \draw (7,0) -- (7,-1) -- (8,-1) -- (8,-1.5) -- (8.5,-1.5) -- (8.5,-2.5) -- (7.5,-2.5) -- (7.5,-1.5) -- (8.5,-1.5);
      \draw (8.5,-1.5) -- (7.5,-2.5);
      \draw (7.5,-1.5) -- (8.5,-2.5);
      \draw (8,-2.5) -- (8,-3) -- (7,-3) node[ground] {} -- (6,-3) to[C, l_=$C_\text{q}$] (6,-1) -- (7,-1);

      \draw (10,0) -- (10,-1) -- (11,-1) to[L, l_=$L_\text{R}$] (11,-3) -- (10,-3) node[ground] {} -- (9,-3) to[C, l_=$C_\text{R}$] (9,-1) -- (10,-1);

      \draw (4,0) -- (4,1) to[C=$C_0$] (10,1) -- (10,0);

      \filldraw (4,0) circle (2pt);
      \filldraw (7,0) circle (2pt);
      \filldraw (10,0) circle (2pt);        
    \end{circuitikz}
  \end{subfigure}

  \medskip

  \begin{subfigure}{0.75\linewidth}
    \begin{tikzpicture}[scale=0.85, transform shape]
      \node[cloud, draw, aspect=2.7, cloud puffs=20] (c1) at (0,0) {\small$\sum_k\omega_k d_k^\dagger d_k$};
      \node[rectangle, draw, minimum width=1cm, minimum height=1cm, thick] (r1) at (3,0) {};
      \node[rectangle, draw, minimum width=1cm, minimum height=1cm, thick] (q1) at (5,0) {};
      \node[rectangle, draw, minimum width=1cm, minimum height=1cm, thick] (r2) at (7,0) {};
      \node[cloud, draw, aspect=2.7, cloud puffs=20] (c2) at (10,0) {\small$\sum_k\omega_k d_k^\dagger d_k$};

      \draw[<->, thick] (c1.east) to [out=30, in=150] node [above] {$\alpha$} (r1.west);
      \draw[<->, thick] (r1.east) to [out=30, in=150] node [above] {$g$} (q1.west);
      \draw[<->, thick] (q1.east) to [out=30, in=150] node [above] {$g$} (r2.west);
      \draw[<->, thick] (r2.east) to [out=30, in=150] node [above] {$\alpha$} (c2.west);
      \draw[<->, thick] (r1.north) to [out=45, in=135] node [above] {$g_{12}$} (r2.north);

      \begin{scope}[scale=0.11, xshift=27.5cm, yshift=-3.5cm]
        \draw[domain=-2.6:2.6, smooth, variable=\x] plot ({\x}, {\x*\x});
        \draw[domain=-1.41:1.41, variable=\x] plot ({\x}, {2});
        \draw[domain=-2:2, variable=\x] plot ({\x}, {4});
        \draw[domain=-2.45:2.45, variable=\x] plot ({\x}, {6});
      \end{scope}

      \begin{scope}[xshift=5cm]
        \draw[domain=-0.4:0.4, variable=\x] plot ({\x}, {-0.25});
        \draw[domain=-0.4:0.4, variable=\x] plot ({\x}, {0.25});
        \draw[<->] (-0.3,0.25) -- (-0.3,-0.25);
        \node at (0.1,0) {$\omega_\text{q}$};
      \end{scope}

      \begin{scope}[scale=0.11, xshift=63.5cm, yshift=-3.5cm]
        \draw[domain=-2.6:2.6, smooth, variable=\x] plot ({\x}, {\x*\x});
        \draw[domain=-1.41:1.41, variable=\x] plot ({\x}, {2});
        \draw[domain=-2:2, variable=\x] plot ({\x}, {4});
        \draw[domain=-2.45:2.45, variable=\x] plot ({\x}, {6});
      \end{scope}
    \end{tikzpicture}

    \medskip
  \end{subfigure}
  \caption{Top: Circuit diagram for the quantum heat valve comprised of a resonator-qubit-resonator system coupled to two resistors on either side. Bottom: An abstraction of the circuit diagram showing the effective nearest-neighbour couplings and the additional direct  phenomenological resonator-resonator coupling not mediated by the qubit. The weak coupling strength $\alpha \ll 1$ of the baths is a dimensionless quantity.}
  \label{fig:Circ-diagram}
\end{figure}

The quantum heat valve, as implemented in Ref.~\cite{Ronzani2018}, consists of a resonator-qubit-resonator setting coupled to two thermal baths at each side of the chain. In the heat valve design, the qubit is installed in between two resonators, which are experimentally realized as coplanar waveguides, and which can  be effectively modelled as quantum harmonic oscillators. These resonators are then coupled to resistive elements, which constitute the thermal baths, i.e., the environment of our open quantum system. The circuit diagram of the system is represented in Fig.~\ref{fig:Circ-diagram}.

Quantizing the system of Fig.~\ref{fig:Circ-diagram} gives its Hamiltonian as
\begin{equation}
  \label{eq:sys-Ham}
  \begin{aligned}
    H_\text{S} = &\Omega_\text{L} a_\text{L}^\dagger a_\text{L} + \Omega_\text{R} a_\text{R}^\dagger a_\text{R} + \omega_\text{q}\sigma_+\sigma_- \\
                 &+ g(a_\text{L}\sigma_+ + a_\text{L}^\dagger\sigma_- + a_\text{R}\sigma_+ + a_\text{R}^\dagger \sigma_-) \\
                 &+ g_{12}(a_\text{L}a_\text{R}^\dagger + a_\text{L}^\dagger a_\text{R})\,,
  \end{aligned}
\end{equation}
where $\Omega_\text{L}, \Omega_\text{R}$ are the resonance frequencies of the left and right resonator respectively and $\omega_\text{q}$ is the frequency of the qubit. The coupling between the resonators and the qubit is equal on both sides and denoted by $g$. Moreover, in line with the model in \cite{Ronzani2018}, we also added a phenomenological direct resonator-resonator coupling denoted by $g_{12}$. The operators $a_\text{L/R}, a^\dagger_\text{L/R}$ are the canonical bosonic annihilation and creation operators respectively, while $\sigma_\pm$ are the corresponding operators for the qubit. In the derivation of the Hamiltonian we have performed the rotating wave approximation, neglecting terms which do not conserve the total number of quanta. We work in units where $\hbar = 1$.

In the language of cQED, the resistors on either side of the chain can be described as an infinite chain of parallel LC-circuits \cite{Vool2017,Cattaneo2021}, equivalent to a series of harmonic oscillators, akin to the Caldeira-Leggett model \cite{Caldeira1983}. The resistors can therefore be quantized to represent an infinite bath of bosonic modes with the Hamiltonian
\begin{equation}
  \label{eq:bath-Ham}
  H_\text{B}^{(b)} = \sum_{k=0}^\infty \omega_{k}^{(b)}d_{k}^{(b)\dagger} d_{k}^{(b)}\,,
\end{equation}
where $\omega_{k}^{(b)}$ is the frequency of the $k$th mode of the left ($b=\text{L}$) or right ($b=\text{R}$) bath. Due to the capacitive connection, the coupling of the bath to the central system is given by a momentum-type coupling as
\begin{equation}
  \label{eq:int-Ham}
  H_\text{I} = \sum_{b=\text{L,R}}\sum_{k=0}^\infty \alpha^{(b)}h_{k}^{(b)}(a^{(b)\dagger} - a^{(b)})(d_k^{(b)\dagger} - d_k^{(b)})\,,
\end{equation}
where $\alpha^{(b)} \ll 1$ is a small, dimensionless parameter quantifying the overall coupling strength to the bath $b$, while $h_k^{(b)}$ is the coupling of the $k$th mode of the (left or right) bath to the system. The parameters $h_k^{(b)}$ are defined via the bath spectral density, which  in its general form is given as
\begin{equation}
  \label{eq:SD-def}
  J^{(b)}(\omega) = \sum_{k=0}^\infty \left|h_k^{(b)}\right|^2\delta(\omega - \omega_k)\,.
\end{equation}
One of the most conventional forms of the spectral density is an ohmic function, which may naturally arise during resistor quantization \cite{Cattaneo2021}. Another common functional form for $J(\omega)$ is a Lorentzian, effectively modelling a filtered ohmic spectral density. This is explained in more detail in Section \ref{sec:modelling-heat-valve} when discussing the phenomenological modelling of the quantum heat valve.

In the quantum heat valve experiment of Ref.~\cite{Ronzani2018}, the qubit was of the transmon type \cite{Koch2007}. Its resonance frequency was tuned by applying an external magnetic flux on the qubit. Following their approach, the qubit resonance frequency is given in terms of the applied flux $\phi$ as \cite{Koch2007}
\begin{equation}
  \label{eq:qubit-freq}
  \omega_\text{q}(\phi) = \sqrt{8E_\text{J}(\phi)E_\text{C}} - E_\text{C}\,,
\end{equation}
where $E_\text{C}$ is the charging energy of the transmon, while the Josephson energy $E_\text{J}(\phi)$ is given by
\begin{equation}
  \label{eq:E_J(phi)}
  E_\text{J}(\phi) = E_{\text{J}0}\abs{\cos(\pi\phi)}\sqrt{1 + d^2\tan^2(\pi\phi)}\,.
\end{equation}
Here the magnetic flux $\phi$ has been normalized by the magnetic flux quantum $\phi_0 = \hbar/(4\pi e)$, becoming a dimensionless quantity. The parameter $d$ describes the critical current asymmetry inside the Josephson junctions of the transmon.

The couplings between the system constituents were capacitive, as indicated in Fig.~\ref{fig:Circ-diagram}, with an additional phenomenological resonator-resonator coupling. For example, it may arise from the stray capacitances within the substrate. The thermal baths, or heat reservoirs, of the open quantum system were realized as normal-metal resistors. 

\subsection{Theoretical model based on Fermi's golden rule and resonator double-counting issue}\label{sec:modelling-heat-valve}

When using the master equation approach in open quantum systems, the heat flow through the system from reservoir $b$ is given by \cite{Alicki1979,Barra2015,Karimi2017}
\begin{equation}
  \label{eq:Heat-flow-me-form}
  P^{(b)} = \Tr[H_\text{S}\mathcal{D}^{(b)}[\rho]]\,,
\end{equation}
where $\mathcal{D}^{(b)}[\rho]$ is the dissipator of the bath $b$ (in our case either left or right) of the master equation acting on the system density matrix $\rho$. This expression is general and  independent of whether the dissipator arises from local or global approaches or from performing the partial or full secular approximations \cite{Cattaneo2019}.

The theoretical model used for fitting to the experimental data of the quantum heat valve in Ref.~\cite{Ronzani2018} is based on a more specific expression \cite{Karimi2017},
\begin{equation}
  \label{eq:heat-flow-sum-expression}
  P^{(\text{L/R})} = \sum_{i,j}\Gamma_{i\to j}^{(\text{L/R})}\rho_{jj}E_{ij}\,,
\end{equation}
where $\Gamma_{i\to j}^{(\text{L/R})}$ is the transition rate from energy level $\epsilon_i$ to $\epsilon_j$ of the system Hamiltonian due to the left or right reservoir, $E_{ij}$ the corresponding energy difference between the eigenenergies and $\rho_{jj}$ is the diagonal element of the density matrix. It can be shown that the physical grounding for this expression is actually given by the global master equation with full secular approximation \cite{Cattaneo2019}, see Appendix \ref{sec:heat-power-deriv}. In other words, Eq.~(\ref{eq:Heat-flow-me-form}) reduces to (\ref{eq:heat-flow-sum-expression}) when the full secular approximation is applied to the master equation.

The transition rates of Eq.~(\ref{eq:heat-flow-sum-expression}) can be obtained for example by using Fermi's golden rule, as done in Ref.~\cite{Ronzani2018}, which corresponds to taking the master equation in the full secular approximation \cite{Alicki1977}. In Ref.~\cite{Ronzani2018} the spectral density of the baths was taken to be of the form of a band-pass filtering Lorentzian
\begin{equation}
  \label{eq:Lorentzian-bath-SD}
  J(\omega) = \chi\frac{\omega}{1 + Q^2(\frac{\omega}{\omega_\text{r}} - \frac{\omega_\text{r}}{\omega})^2}\,,
\end{equation}
where the parameters $\chi$ and $Q$ modify the height and width of the distribution respectively and $\omega_\text{r}$ sets the position of the maximum. Practically this kind of situation corresponds to a resonator of frequency $\omega_\text{r}$ being incorporated into the environment, such that it modifies the spectral density of the bath \cite{Cattaneo2021}. In Ref.~\cite{Ronzani2018} the situation is exactly this, as in the heat valve modelling the frequency $\omega_\text{r}$ is taken to be the resonance frequency of the resonator, i.e., in our case $\omega_\text{r} = \Omega_\text{L/R}$, while $Q$ represents the quality factor of the cavity.

Incorporating the resonator into the environment via modification of the spectral density can be justified in the case of strong bath-resonator coupling, while the coupling to the rest of the system from the resonator remains weak. This procedure was applied in \cite{Karimi2016} and experimentally verified in the \textit{non-Hamiltonian} case of Ref.~\cite{Ronzani2018}. However, the key point to understand is that in this situation the Hamiltonian of the open quantum system of interest should not include resonator terms. In the case of the quantum heat valve, the Hamiltonian would then be simply that of the qubit alone.

However, if the bath-resonator coupling is also taken to be weak, as in the \textit{quasi-Hamiltonian} model of Ref.~\cite{Ronzani2018}, incorporating the resonator to the environment raises a question concerning the physical interpretation of the situation. Since the resonators are already taken into account in the Hamiltonian when computing the heat flow using Eq.~(\ref{eq:heat-flow-sum-expression}) in Ref.~\cite{Ronzani2018}, how can resonators also be incorporated into the bath? This creates a resonator double counting issue, which we aim to solve using the global ME PSA approach discussed later in Section \ref{sec:results}, avoiding the usage of Lorenzian baths.

A more advanced theoretical model for the quantum heat valve was discussed in \cite{Xu2021}. There, the non-equilibrium steady state heat flow is solved exactly using hierarchical equations of motion with different bath spectral densities, both structured and ohmic with a Debye cutoff. However, as the HEOM method is computationally intensive, it is interesting to see whether correct results can be obtained using a more light weight, albeit perturbative, method like the master equation approach, while still retaining the physically meaningful description for the baths being of ohmic nature. This is discussed in more detail in Section \ref{sec:results}.


\section{Modelling the quantum heat valve using a global master equation in partial secular approximation}\label{sec:results}

\subsection{Theoretical model}
Our approach relies on describing the system as weakly coupled to the thermal baths on either side of the qubit-resonator chain. The baths are assumed to be ohmic, with a spectral density of the form  
\begin{equation}
  J(\omega) = \chi\frac{\omega}{1 + \omega^2/\omega_\text{c}^2}\,,
\end{equation}
where $\chi$ is a coefficient describing the coupling strength of the individual bath modes to the system and $\omega_\text{c}$ is a cutoff frequency. This choice is justified by the baths being resistive elements, see for instance the theoretical model in \cite{Cattaneo2021}. 

The dynamics of the quantum system are in general driven by the Redfield equation \cite{Cattaneo2019,Vaaranta2026}
\begin{equation}
  \label{eq:Redfield}
  \begin{aligned}
    \dot{\rho}(t) = &-\i[H_\text{S} + H_\text{LS}, \rho(t)] \\
                    &+ \sum_{b=\text{L,R}}\left(\alpha^{(b)}\right)^2\sum_{\omega, \omega^\prime}\gamma^{(b)}(\omega, \omega^\prime)\left(A(\omega)\rho(t)A^\dagger(\omega^\prime) - \frac{1}{2}\{A^\dagger(\omega^\prime)A(\omega), \rho\}\right)\,,
  \end{aligned}
\end{equation}
where $H_\text{LS}$ is the ``Lamb-shift Hamiltonian''(see \cite{Cattaneo2019,Vaaranta2026} for further details) 
and $\alpha^{(b)}\ll 1$ is a weak dimensionless system-bath coupling. The quantity $\gamma^{(b)}(\omega, \omega^\prime)$ is given by
\begin{equation}
  \gamma^{(b)}(\omega, \omega^\prime) = \Gamma^{(b)}(\omega) + \Gamma^{(b)*}(\omega^\prime)\,,
\end{equation}
where
\begin{equation}
  \Gamma^{(b)}(\omega) = \int_0 ^\infty\d s\left\langle\tilde{B}^{(b)\dagger}(s)\tilde{B}^{(b)}(0)\right\rangle_{\rho_\text{B}}\e^{\i\omega s}
\end{equation}
is the one-sided Fourier transform of the bath correlation function over the state of the bath $\rho_B$, where $B^{(b)}$ denote the bath operators that couple to the system operators in the interaction Hamiltonian.\footnote{As the interaction Hamiltonian is given by Eq.~(\ref{eq:int-Ham}), the bath operator $B^{(b)}$ is in our case given by $B^{(b)} = d_k^{(b)\dagger} - d_k^{(b)}$.} The jump operators in Eq.~(\ref{eq:Redfield}) are defined as
\begin{equation}
  \label{eq:jump-op-def}
  A(\omega) = \sum_{\epsilon_j-\epsilon_i = \omega} \left\langle\epsilon_i|A|\epsilon_j\right\rangle\ket{\epsilon_i}\bra{\epsilon_j}\,,
\end{equation}
where $\ket{\epsilon_i}$ is an eigenstate of the full system Hamiltonian in Eq.~(\ref{eq:sys-Ham}).

\begin{figure}
  \centering
  \includegraphics[width=0.75\linewidth]{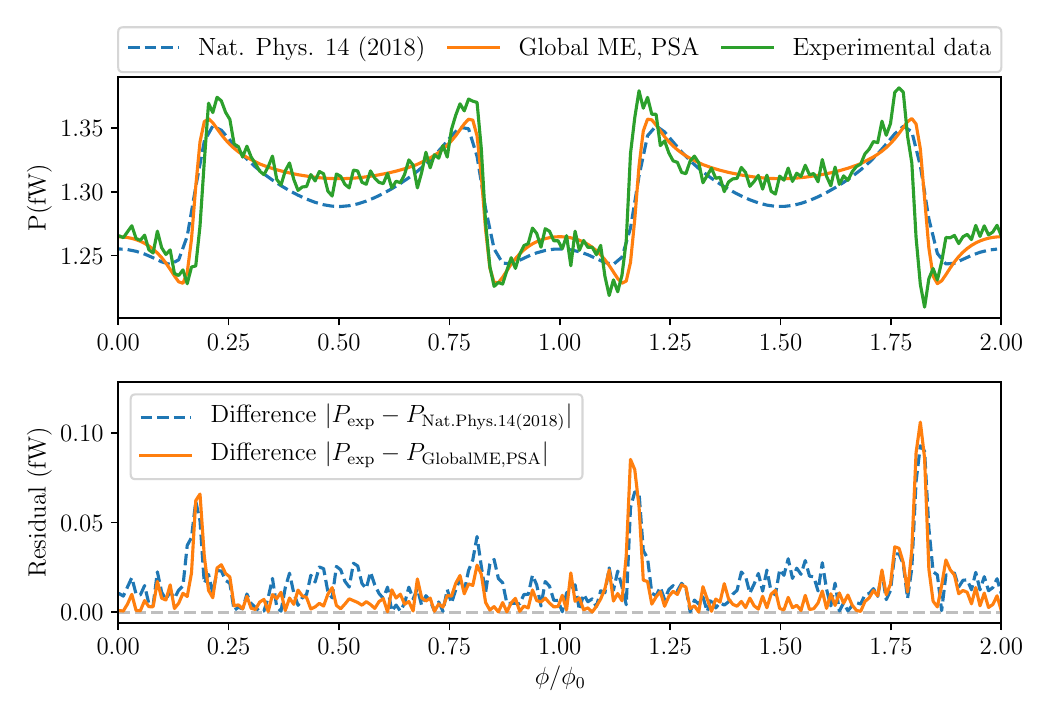}
  \caption{The numerically fitted theoretical model (dashed blue line), from Ref.~\cite{Ronzani2018}, is compared against the the master equation approach with partial secular approximation (solid orange line) and the experimental data (solid green line). We see that the model in \cite{Ronzani2018} matches the experimental data well but fails, to some extent, to capture the correct curvature of the data around half and integer flux. This behaviour is improved in the global ME PSA approach. On the bottom figure the mismatch between the experimental data and both models is plotted. The parameters shared by both models are $\Omega_\text{L} = \Omega_\text{R} = 2\pi\cdot5.3122\,\text{GHz}\,, E_\text{C}/\Omega_\text{L} = 0.15\,, d = 0.385\,, T_\text{L} = 308\,\text{mK}\,,T_\text{R} = 100\,\text{mK}$. Additionally, the  model from \cite{Ronzani2018} uses $Q = 20\,, g/\Omega_\text{L} = 0.0171\,, g_{12}/\Omega_\text{L} = -0.0217\,,E_{\text{J}0}/\Omega_\text{L} = 27.54$, while the global ME PSA approach uses $g/\Omega_\text{L} = 0.015\,, g_{12}/\Omega_\text{L} = 0.007\,, E_{\text{J}0}/\Omega_\text{L} = 28.75$ and $\alpha^{(L)} = \alpha^{(R)} = 0.04$. Moreover, in the global ME PSA approach we have set the cutoff frequency to $\omega_\text{c}/\Omega_\text{L} = 50$, the partial secular approximation cutoff $C_\text{PSA} = 100$ and the resonators have been truncated to three levels.\protect\footnotemark}
  \label{fig:QHV_og_heatflow_model}
\end{figure}
\footnotetext{The cutoff for the partial secular approximation is a "confidence threshold" for neglecting fast rotating terms of the master equation dissipator via observing whether the inequality $C_\text{PSA}|\omega - \omega^\prime|^{-1} < \tau_\text{R}$ is satisfied. Here $\tau_\text{R} = \alpha^{-2}|\Gamma(\omega)|^{-1}$ is the estimated relaxation time scale of the system. See Ref.~\cite{Vaaranta2026} for more details.}

It should be noted that we have not applied the secular approximation to Eq.~\eqref{eq:Redfield}; we are keeping all cross terms with $\omega\neq\omega^\prime$ in order to obtain more accurate results. Indeed, the full secular equation may neglect slowly rotating cross terms, leading to incorrect physical predictions \cite{Cattaneo2019}. At the same time, while the master equation in full secular approximation is always in Lindblad form, it is known that the Redfield equation is not guaranteed to be completely positive. However, if the Born-Markov approximations leading to the Redfield equation are justified, that is, if $\alpha$ is sufficiently weak and the autocorrelation functions of the environment decay sufficiently fast \cite{Cattaneo2019,Vaaranta2026},  eventual violations of positivity preservation by the dynamics generated by \eqref{eq:Redfield} are usually negligible \cite{Vaaranta2026,Hartmann2020}.  

\subsection{Comparison with Fermi's golden rule and experimental data}

As noted earlier, if we then compute the heat flow using the global master equation with the full secular approximation, we obtain the heat flow given by Eq.~(\ref{eq:heat-flow-sum-expression}), as done in Ref.~\cite{Ronzani2018}, which arises from direct application of Eq.~(\ref{eq:Heat-flow-me-form}) (see Appendix \ref{sec:heat-power-deriv}). In Fig.~\ref{fig:QHV_og_heatflow_model} we plot the heat flow given by Eq.~(\ref{eq:heat-flow-sum-expression}) and compare it with both the experimental data of Ref.~\cite{Ronzani2018} and our approach based on the careful application of the partial secular approximation, which removes only the fast-rotating cross terms in Eq.~\eqref{eq:Redfield} \cite{Vaaranta2026,Cattaneo2019}.

In the case of the theoretical model from Ref.~\cite{Ronzani2018} (dashed blue line), the heat flow is computed numerically by fitting the model to the experimental data. It is important to note that, in such a complex experiment, most circuit parameters cannot be determined independently with sufficient precision and must therefore be inferred through a suitable fit to the theoretical model. Analytical expressions for the decay rates and heat flow as functions of the circuit parameters are provided in the supplementary material of Ref.~\cite{Ronzani2018}. A numerical fit is thus justified, provided that the resulting parameters lie within physically reasonable ranges, which is indeed the case in Ref.~\cite{Ronzani2018}.

From the fit we obtain values for the circuit parameters and  for the spurious  resonator-resonator capacitance that is related to $g_{12}$ (see Fig.~\ref{fig:Circ-diagram}), which we consider as a model-dependent phenomenological parameter. From Fig.~\ref{fig:QHV_og_heatflow_model}, we observe that the simple model of the resonator-qubit-resonator system coupled to the two thermal baths captures the non-equilibrium steady state heat flow rather well. In this case the baths have a structured spectral density with a Lorenzian peak at the resonance frequency of the resonator.

Next, we implement the partial secular equation numerically using a code recently developed by one of us \cite{Vaaranta2026,Vaaranta2025_GitHub}. We compute the non-equilibrium steady state heat flow through the system and compare our results with the experimental data. Notably, in this case we assume the baths to be ohmic, with spectral density of the form in Eq.~(\ref{eq:Lorentzian-bath-SD}), rather than structured with Lorenzian peaks. The results are depicted through the solid orange line in Fig.~\ref{fig:QHV_og_heatflow_model}. 

In our model, we use the same circuit parameters obtained from the fit to the theoretical model of Ref.~\cite{Ronzani2018}, while slightly tuning some of them to better reproduce the experimental curves. This procedure is justified for the same reasons as discussed above: since these parameters cannot be precisely determined a priori, it is reasonable to fit them, provided that the resulting values remain within physically reasonable ranges.

More specifically, the differences between the two sets of parameters are minimal. The most notable change concerns the resonator–resonator coupling, which changes from $g_{12}/\Omega_\text{L} = -0.0217$ to $g_{12}/\Omega_\text{L} = 0.007$ between the models. However, this is not worrying as this parameter captures a spurious coupling between the two resonators that was not originally designed in the experiment and its value cannot be precisely determined, nor can its sign. This is taken as a model-dependent parameter. Additionally, we changed the qubit-resonator coupling slightly, from $g/\Omega_\text{L} = 0.0171$ to $g/\Omega_\text{L} = 0.015$. However, this change, being around $10\%$, is within the experimental uncertainty of the original design parameters.

From Fig.~\ref{fig:QHV_og_heatflow_model} we observe that using the ohmic spectral density and partial secular global master equation (ME PSA), we are able to obtain results that are as well, or even better, matching the experimental data as in the more simplified formalism. Especially the sharp features in the data are captured better in this case than by the model from Ref.~\cite{Ronzani2018}. In addition, all parameters have a well-defined physical meaning, as they arise directly from the microscopic derivation of the model, which is now conceptually well grounded as we have removed the resonator double-counting issue. 
Therefore, we argue that the computationally expensive HEOM method of Ref.~\cite{Xu2021} is not necessarily needed in order to obtain accurate results for the heat flow through the quantum heat valve in the non-equilibrium steady state.

From our analysis, we find that when the baths are ohmic (i.e., in the absence of a resonator double-counting issue), the global master equation with the \textit{full} secular approximation fails to capture the heat flow through the system even qualitatively (related plots are omitted for brevity). As a consequence, we conclude that the naive application of the full secular approximation requires the unphysical assumption of Lorentzian baths, in addition to the numerical fit that takes advantage of the complexity of this spectral density function, in order to reproduce the heat flow behaviour correctly. Moreover, since the full secular approach is equivalent to the Fermi’s golden rule method, with the heat flow given by Eq.~(\ref{eq:heat-flow-sum-expression}) and depending only on the diagonal elements of the steady-state density matrix, we conclude that steady-state coherences clearly play a crucial role in the physically realistic modeling of heat flow through the quantum heat valve.

\subsection{Unified master equation}

\begin{figure}
  \centering
  \includegraphics[width=0.75\linewidth]{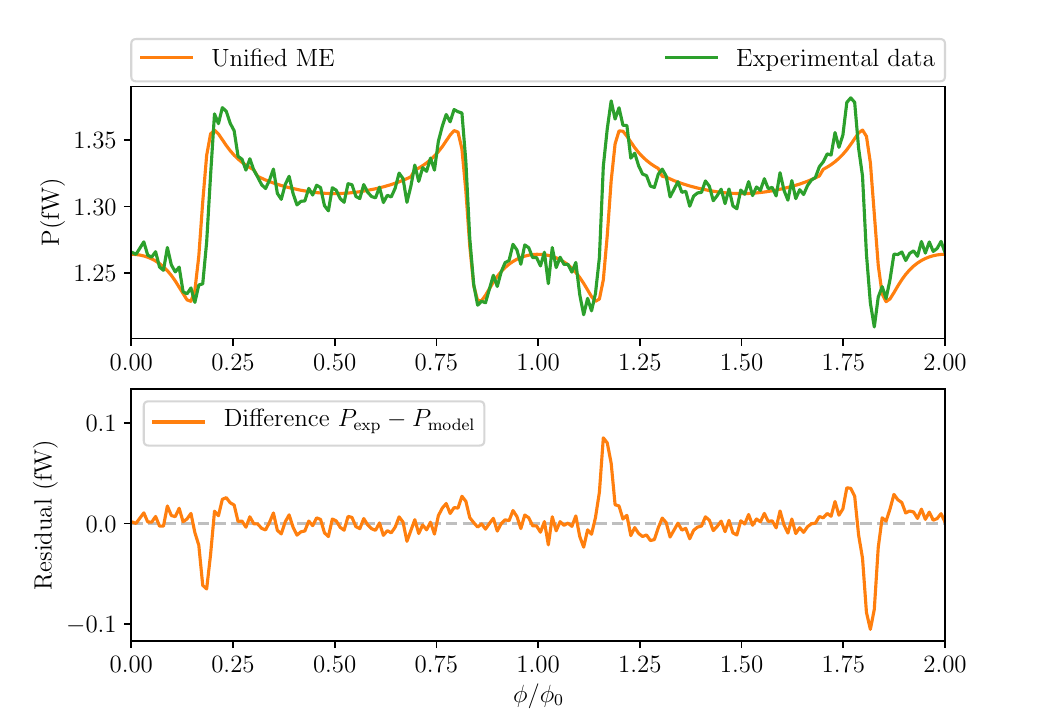}
  \caption{Non-equilibrium steady state heat flow computed using the unified master equation approach and comparison with experimental data. We can see that also the unified approach is able to capture the experimental behaviour very well, essentially numerically overlapping the more conventional global ME PSA approach from Fig.~\ref{fig:QHV_og_heatflow_model}. The parameters used in the unified approach are the same as in Fig.~\ref{fig:QHV_og_heatflow_model} for the global ME PSA approach.}
  \label{fig:QHV_global_psa_heatflow_unified}
\end{figure}

To further explore the use of perturbative master equations to describe the experimental behaviour, we also compute the non-equilibrium steady state heat flow using the \textit{unified master equation} \cite{Trushechkin2021}. The unified equation is a modification of the Redfield equation in partial secular approximation that guarantees complete positivity and the Lindblad form. In this approach the Bohr frequencies are grouped into distinct clusters, each of which is associated with a single jump frequency and a cluster jump operator. For further details we refer the readers to the original paper \cite{Trushechkin2021} and to Ref.~\cite{Potts2021}. 

We computed the unified master equation through the same code we used for the partial secular approximation \cite{Vaaranta2026}. It is worth noting that the frequency clustering of the unified approach reduces the number of jump operators significantly, making this method numerically more lightweight, while still capturing the correct behaviour of the open quantum system dynamics.

 The result from the unified master equation approach are shown in Fig.~\ref{fig:QHV_global_psa_heatflow_unified}, where the used parameters are the same as in Fig.~\ref{fig:QHV_og_heatflow_model} for the global ME PSA approach. We observe that the theoretical model agrees well with experimental data and essentially perfectly mimics the behaviour obtained through the global ME PSA approach. At the same time, we stress that the unified approach is computationally \textit{ten times faster} than the partial secular one.

\section{Conclusion and outlook}\label{sec:conclusion}

In this paper we have provided a theoretical model for the experimental results on the non-equilibrium steady-state heat flow in the quantum heat valve presented in Ref.~\cite{Ronzani2018}. Our model is based on a master equation in partial secular approximation, contrary to the full secular one employed in \cite{Ronzani2018}. We have shown that a careful application of the partial secular approximation on the Redfield master equation is able to provide an accurate description of the experimental data. This approach also assigns a clear physical meaning to all the parameters of the theoretical model. Moreover, we have observed that similar predictions can be made also through the use of the unified master equation \cite{Trushechkin2021}, corroborating the validity of perturbative master equations for describing the experimental data.

In our approach,  the resonator is part of the central open quantum system of interest. Moreover, the two thermal baths coupled to the central resonator–qubit–resonator chain are modelled with ohmic spectral density. This is in contrast to the original model from Ref.~\cite{Ronzani2018}, where the spectral density of the baths were modified with a Lorentzian structure peaked at the resonance frequency of the resonator. As the resonator was also considered as a part of the central quantum system of interest, this created a resonator double counting issue.

The perturbative master equation approach with partial secular approximation is able to capture the relevant behaviour of the heat flow even more accurately than the original approach \cite{Ronzani2018}, while being conceptually consistent (keep in mind, however, that in both cases the precise circuit parameters arise from a fit of the theory to the experiment). Indeed, it avoids the double counting of the resonator and it does not discard the slowly rotating terms in the dissipator of the master equation that are incorrectly neglected in the full secular approach. The HEOM approach \cite{Xu2021} is also accurate in describing the experimental data but is much more  computationally challenging. The power of the partial secular/unified approach is its relative simplicity both conceptually and in the numerical implementation.

The question of how to treat coupling of noise, or an external ac signal at a given frequency on a hybrid quantum system occurs frequently in modern quantum technologies. Here we focused on a superconducting system consisting of a transmon qubit and two resonators, subject to capacitively coupled noise sources formed of thermal baths that are nanoscale resistive normal metal elements. The potential double-counting issue arises more generally, for instance in a recent work on a superconducting flux qubit coupled inductively (galvanically) to resonators \cite{Upadhyay2025}. The key issue in the treatment is the hierarchy of relevant couplings in each setup: for example, if the coupling between the resonators and qubit in the presented system is much weaker than that between a resonator and the bath, the double-counting plays no role since the resonator can be absorbed as part of the environment in the model. In the opposite regime, the situation is different and the model presented here makes a difference. It is worth pointing out that under some conditions, a Caldeira-Leggett type description \cite{Caldeira1983} of the full system including the (finite size) heat baths is another option to treat the problem \cite{Pekola2022,Mäkinen2025,DoGoMG2020}, at least for benchmarking purposes.

\backmatter

\bmhead{Acknowledgements}
The authors wish to thank Giuseppe Falci for useful discussions. We acknowledge the financial support of the Research Council of Finland through the Finnish Quantum Flagship project (358878, UH) and (358877, Aalto). MC acknowledges funding from COQUSY project PID2022-140506NB-C21 funded by MCIN/AEI/10.13039/501100011033.



\begin{appendices}

  \section{Heating power derivation}\label{sec:heat-power-deriv}
  The equation for the heating power, Eq.~\eqref{eq:heat-flow-sum-expression}, can be intuitively understood by connecting the populations of the energy levels to the decay rates of the master equation in the full secular approximation. In this Appendix we show the equivalence between Eqs.~(\ref{eq:Heat-flow-me-form}) and (\ref{eq:heat-flow-sum-expression}), which reflects the equivalence between Fermi's golden rule and the global master equation with full secular approximation \cite{Alicki1977}. 

  We begin with the definition for the heating power from one bath to another \cite{Alicki1979,Barra2015}
  \begin{equation}
    P_b = \Tr[H_\text{S}\mathcal{D}_b[\rho_\infty]]\,,
  \end{equation}
  where $\mathcal{D}_b[\rho_\infty]$ is the dissipator of the bath $b$ of the master equation acting on the steady state of the system $\rho_\infty$, defined by $\dot{\rho}_\infty = 0$,  which corresponds to the zero-eigenvalue eigenstate of the Liouvillian superoperator. The dissipator can in general be written as
  \begin{equation}
    \label{eq:dissipator}
    \mathcal{D}_b[\rho] = \sum_{\omega, \omega^\prime} \gamma_b(\omega, \omega^\prime)\left(A(\omega)\rho A^\dagger(\omega^\prime) - \frac{1}{2}\left\{A^\dagger(\omega^\prime)A(\omega), \rho\right\}\right)\,.
  \end{equation}
  Here $\gamma_b(\omega, \omega^\prime)$ is a rate that characterizes the strength of the dissipation in the system due to bath $b$ and $A(\omega)$ are the jump operators defined as
  \begin{equation}
    \label{eq:jump-op-def}
    A(\omega) = \sum_{\epsilon_j-\epsilon_i = \omega} \langle\epsilon_i|A|\epsilon_j\rangle\ket{\epsilon_i}\bra{\epsilon_j}\,,
  \end{equation}
  with the states $\ket{\epsilon_i}$ being the eigenstates of the Hamiltonian as $H_\text{S}\ket{\epsilon_i} = \epsilon_i\ket{\epsilon_i}$.

  Let us now consider the dissipator in Eq.~(\ref{eq:dissipator}) in the full secular approximation, which neglects all the terms of the dissipator with unequal frequencies $\omega \neq \omega^\prime$. Performing the full secular approximation and using the definition (\ref{eq:jump-op-def}) for the jump operators transforms Eq.~(\ref{eq:dissipator}) into
  \begin{multline}
    \mathcal{D}_b(\rho) = \sum_\omega\sum_{\substack{\epsilon_j-\epsilon_i = \omega\,, \\ \epsilon_k - \epsilon_l = \omega}} \gamma_b(\omega)\bigg[\langle\epsilon_i|A|\epsilon_j\rangle\langle\epsilon_k|A|\epsilon_l\rangle\langle\epsilon_j|\rho|\epsilon_k\rangle\ket{\epsilon_i}\bra{\epsilon_l} \\
    - \frac{1}{2}\delta_{\epsilon_i\epsilon_l}\left\{\langle\epsilon_k|A|\epsilon_l\rangle\langle\epsilon_i|A|\epsilon_j\rangle\ket{\epsilon_k}\bra{\epsilon_j},\rho\right\}\bigg]\,.
  \end{multline}
  Next we can act upon this expression with the Hamiltonian, remembering that $\ket{\epsilon_i}$ is an eigenstate of the Hamiltonian. Then we can take the trace and apply its linearity and cyclicity properties to obtain the following expression:
  \begin{equation}
    \Tr[H_\text{S}\mathcal{D}_b[\rho]] = \sum_\omega\sum_{\substack{\epsilon_j-\epsilon_i = \omega\,, \\ \epsilon_k - \epsilon_l = \omega}}\gamma_b(\omega)\langle\epsilon_i|A|\epsilon_j\rangle\langle\epsilon_k|A|\epsilon_l\rangle\langle\epsilon_j|\rho|\epsilon_k\rangle\left(\epsilon_i - \frac{\epsilon_k}{2} - \frac{\epsilon_j}{2}\right)\delta_{\epsilon_i\epsilon_l}\,.
  \end{equation}
  The Kronecker delta sets $\epsilon_l = \epsilon_i$ in the second summation. However, due to the condition of the two sums, we also get $\epsilon_j - \epsilon_i = \epsilon_k - \epsilon_i \Rightarrow \epsilon_k = \epsilon_j$. Therefore the expression simplifies to
  \begin{equation}
    \Tr[H_\text{S}\mathcal{D}_b[\rho]] = \sum_\omega\sum_{\epsilon_j - \epsilon_i = \omega}\gamma_b(\omega)\langle\epsilon_i|A|\epsilon_j\rangle\langle\epsilon_j|A|\epsilon_i\rangle\langle\epsilon_j|\rho|\epsilon_j\rangle(\epsilon_i - \epsilon_j)\,.
  \end{equation}
  We can therefore notice that the trace equation [Eq.~(\ref{eq:Heat-flow-me-form})] becomes heat flow equation in the sum form [Eq.~(\ref{eq:heat-flow-sum-expression})] when the full secular approximation has been applied to the master equation:
  \begin{equation}
    P_b = \Tr[H_\text{S}\mathcal{D}_b[\rho]] = \sum_{i,j}\Gamma_{i\to j, b}\rho_{jj}E_{ij}\,,
  \end{equation}
  where the transition rate is defined as $\Gamma_{i\to j, b} = \gamma_b(\omega)\langle\epsilon_i|A|\epsilon_j\rangle\langle\epsilon_j|A|\epsilon_i\rangle$, the diagonal elements of the density matrix are given as $\rho_{jj} = \langle\epsilon_j|\rho|\epsilon_j\rangle$ and the transition energy as $E_{ij} = \epsilon_i - \epsilon_j$.
  
\end{appendices}


\bibliography{library}

\end{document}